\documentclass[10pt,conference]{IEEEtran}

\usepackage{amsmath,amssymb,amsthm} 
\usepackage[dvips]{graphicx}
\usepackage{cite}
\usepackage{balance}
 
\newtheorem{example}{Example}
\newtheorem{remark}{Remark}
\newtheorem{proposition}{Proposition}
\newtheorem{theorem}{Theorem}

\begin{document}

\title{Improvement of BP-Based CDMA Multiuser Detection by Spatial Coupling}

\author{
  \IEEEauthorblockN{Keigo Takeuchi}
  \IEEEauthorblockA{
    Dept.\ Commun.\ Engineering \& Inf.\\ 
    University of Electro-Communications\\
    Tokyo 182-8585, Japan\\
    Email: takeuchi@ice.uec.ac.jp
  }
  \and
  \IEEEauthorblockN{Toshiyuki Tanaka}
  \IEEEauthorblockA{
    Graduate School of Informatics \\ 
    Kyoto University \\ 
    Kyoto 606-8501, Japan \\
    Email: tt@i.kyoto-u.ac.jp
  }
  \and
  \IEEEauthorblockN{Tsutomu Kawabata}
  \IEEEauthorblockA{
    Dept.\ Commun.\ Engineering \& Inf.\\ 
    University of Electro-Communications\\
    Tokyo 182-8585, Japan\\
    Email: kawabata@ice.uec.ac.jp
  }
}


\maketitle

\begin{abstract}
Kudekar et al.\ proved that the belief-propagation (BP) threshold for 
low-density parity-check codes can be boosted up to the 
maximum-a-posteriori (MAP) threshold by spatial coupling. In this paper, 
spatial coupling is applied to randomly-spread code-division multiple-access 
(CDMA) systems in order to improve the performance of BP-based multiuser 
detection (MUD). Spatially-coupled CDMA systems can be regarded as multi-code 
CDMA systems with two transmission phases. 
The large-system analysis shows that spatial coupling can 
improve the BP performance, while there is a gap between the BP performance 
and the individually-optimal (IO) performance.  
\end{abstract}

\section{Introduction}
The belief-propagation (BP) threshold of a low-density parity-check (LDPC) 
convolutional code~\cite{Felstrom99} has been shown to coincide with the 
maximum-a-posteriori (MAP) threshold of the corresponding LDPC block 
code~\cite{Kudekar10,Lentmaier10}. Since LDPC convolutional codes can be 
regarded as a spatially-coupled chain of LDPC block codes, this phenomenon is 
referred to as threshold saturation via spatial coupling~\cite{Kudekar10}. 

Recently, we proposed a phenomenological model for explaining this 
phenomenon~\cite{Takeuchi111}. In the phenomenological study, spatial coupling 
is regarded as a general method for conveying an suboptimal solution to 
the optimal solution. In this paper, we apply the principle of spatial 
coupling to code-division multiple-access (CDMA) systems in order to improve 
the performance of BP-based multiuser detection (MUD). 

It is important in CDMA systems to mitigate multiple-access 
interference (MAI), by using MUD~\cite{Verdu98}. 
Let us consider conventional $K$-user randomly-spread CDMA systems with 
spreading factor~$N$. 
Tanaka~\cite{Tanaka02} used the replica method to analyze the performance of 
the optimal MUD, called the individually-optimal (IO) 
receiver~\cite{Verdu98}, in the large-system limit, where $K$ and $N$ tend to 
infinity while the system load $\beta=K/N$ is kept constant. 
For $\beta<\beta_{\mathrm{IO}}$, with $\beta_{\mathrm{IO}}$ denoting a 
critical system load, the mean-squared error (MSE) for soft-decisions of 
the IO receiver takes a small value. Thus, the IO receiver can mitigate MAI 
successfully for 
$\beta<\beta_{\mathrm{IO}}$. For $\beta>\beta_{\mathrm{IO}}$, on the other 
hand, the MSE for the IO receiver takes a large value. This result implies 
that MAI cannot be mitigated for $\beta>\beta_{\mathrm{IO}}$. 
Thus, $\beta_{\mathrm{IO}}$ can be regarded as a threshold between the 
MAI-limited region and the non-limited region. Since the system load $\beta$ 
is proportional to the sum rate, the IO threshold 
$\beta_{\mathrm{IO}}$ provides a performance index for the IO receiver.  

The computational complexity of the IO receiver grows exponentially in the 
number of users. Kabashima~\cite{Kabashima03} proposed a low-complexity 
iterative MUD algorithm based on BP. It was shown numerically that the 
algorithm converges for large systems. Furthermore, the BP-based receiver for 
large systems can achieve the MSE of the IO receiver for 
$\beta<\beta_{\mathrm{BP}}$, with a system load
$\beta_{\mathrm{BP}}(<\beta_{\mathrm{IO}})$, 
while the MSE takes a larger value for $\beta>\beta_{\mathrm{BP}}$.  
Thus, the BP threshold $\beta_{\mathrm{BP}}$ can be regarded as another 
critical threshold between the MAI-limited and non-limited regions for the 
BP-based receiver. See \cite{Montanari06,Guo08} for a rigorous treatment 
based on sparsely-spread CDMA systems. 

\begin{figure}[t]
\begin{center}
\includegraphics[width=\hsize]{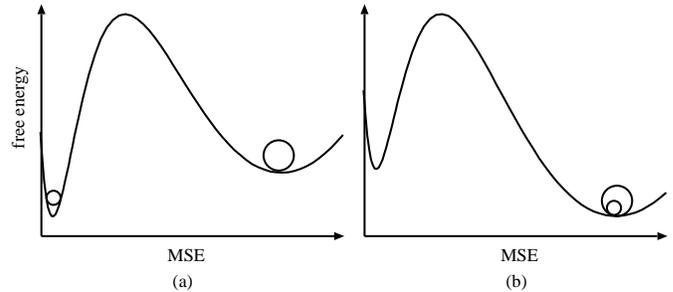}
\end{center}
\caption{
Typical landscape of free energy as a function of MSE. (a) 
$\beta\in(\beta_{\mathrm{BP}},\beta_{\mathrm{IO}})$. 
(b) $\beta>\beta_{\mathrm{IO}}$. The large (small) balls represent the 
solutions for the BP-based receiver (the IO receiver).  
}
\label{fig1} 
\end{figure}

What is occurring for $\beta$ between $\beta_{\mathrm{BP}}$ and 
$\beta_{\mathrm{IO}}$? The MSE for the IO receiver is given by the global 
stable solution of a potential energy function, called free 
energy~\cite{Tanaka02}. On the other hand, the MSE for the BP-based 
receiver is given by the stable solution of the free energy corresponding to 
the largest MSE~\cite{Kabashima03} (See also \cite[Fig.~2(a)]{Takeda06}). 
This solution is metastable for $\beta\in
(\beta_{\mathrm{BP}},\beta_{\mathrm{IO}})$, as shown in Fig.~\ref{fig1}(a), 
while it is the global stable solution for $\beta>\beta_{\mathrm{IO}}$. 
Thus, the MSE for the BP-based receiver is trapped in a metastable solution 
for $\beta\in(\beta_{\mathrm{BP}},\beta_{\mathrm{IO}})$, while the MSE 
for the IO receiver approaches the global stable solution. 

Instead of constructing a BP-based receiver for spatially-coupled CDMA 
systems, in this paper, we derive the BP fixed-point equation, 
characterizing the performance of the BP-based receiver after sufficiently 
many iterations, by using the replica method. 
We will show that the MSE for BP-based receivers can be brought to the 
global stable solution of an effective potential energy function by spatial 
coupling, i.e., the BP threshold is boosted up to a threshold. Unfortunately, 
the threshold does not coincide with the IO threshold, because the potential 
energy is different from the free energy for conventional CDMA systems. 
The rest of this paper is organized as follows: 
In Section~\ref{sec2} we review our phenomenological study~\cite{Takeuchi111}. 
In Section~\ref{sec3} we define spatially-coupled CDMA systems. 
Section~\ref{sec4} presents a large-system analysis of the 
spatially-coupled CDMA systems. In Section~\ref{sec5} we show that 
spatial coupling can improve the performance of BP-based MUD. 
Section~\ref{sec6} concludes this paper. 

\section{A Phenomenological Study} \label{sec2} 
Density evolution (DE) is a powerful method for analyzing the performance 
of BP-based algorithms for graphical models~\cite{Richardson08}. In the 
BP-based algorithms, messages are iteratively exchanged between variable nodes 
and function nodes on the factor graph. 
The densities of the messages are characterized by a few macroscopic 
parameters, such as the average error probability for LDPC codes over binary 
erasure channel (BEC)~\cite{Kudekar10} or the MSE for conventional 
CDMA systems~\cite{Guo08}. DE is a method for deriving a closed-form 
time-evolution equation for the macroscopic parameters, called the DE equation. 
For simplicity, we consider the case in which each density is described by 
one macroscopic parameter. 

Let us consider a spatial coupling of $L$ subsystems. The densities of the 
messages for the spatially-coupled system in iteration~$i$ are characterized 
by macroscopic parameters $\{y_{i}(x=l/L)\}$ at spatial positions 
$l=0,\ldots,L-1$. We assume that the DE equation for the macroscopic 
parameters in $L\rightarrow\infty$ is given by a dynamical system on 
$x\in[0,1]$ with a {\em small} diffusion coefficient $0<D\ll 1$, 
\begin{equation} \label{dynamical_system} 
y_{i+1}(x) - y_{i}(x) = - \frac{dU}{dy}(y_{i}(x);\beta) 
+ D\frac{d^{2} y_{i}}{dx^{2}}, 
\end{equation}
with $U(y;\beta)$ denoting a potential energy function, in which $\beta$ is a 
parameter characterizing the subsystem, such as the system load for CDMA 
systems. The potential energy is assumed to have two stable solutions in a 
region of $\beta$. Note that the potential energy depends on the 
position~$x$ only through $y_{i}(x)$. 

Let $y_{\min}$ denote the global stable solution of the potential 
energy $U(y;\beta)$. We impose the boundary conditions 
$y_{i}(0)=y_{i}(1)=y_{\min}$. 
This type of boundary corresponds to termination for LDPC convolutional codes. 
One may expect that the information about the global stable solution at the 
boundaries diffuses over the whole system, because the diffusion term in 
(\ref{dynamical_system}) smooths $y_{i}(x)$ spatially.  
 In fact, this intuition is correct. 
\begin{theorem}[Takeuchi et al.~\cite{Takeuchi111}] \label{th1} 
Suppose that $y_{i}(x)$ converges to a stationary solution $y(x)$ 
in $i\rightarrow\infty$, satisfying 
\begin{equation} \label{stationary_system} 
0 = - \frac{dU}{dy}(y(x);\beta) + D\frac{d^{2} y}{dx^{2}}. 
\end{equation}
If $y_{\min}$ is the unique minimizer of the potential energy $U(y;\beta)$, 
then,  the uniform solution $y(x)=y_{\min}$ is the unique stationary 
solution to (\ref{dynamical_system}). 
\end{theorem}  

What occurs when the boundaries are fixed to a metastable solution of the 
potential energy $U(y;\beta)$? Numerical simulations~\cite{Takeuchi111} 
imply that $y_{i}(x)$ converges to a spatially-nonuniform stationary 
solution for sufficiently small $D>0$. Let us define the BP threshold 
$\beta_{\mathrm{BP}}^{(\mathrm{SC})}$ for the spatially-coupled system 
as the supremum of $\beta$ such that $y_{i}(x)$ converges to the uniform 
solution $y(x)=y_{\min}$ in $i\rightarrow\infty$. Theorem~\ref{th1} and 
the observation described just above imply that  
$\beta_{\mathrm{BP}}^{(\mathrm{SC})}$ is given by the point $\beta$ at which 
the potential energy $U(y;\beta)$ at one stable solution $y=y_{\min}$ is 
equal to that at the other stable solution $y=\tilde{y}_{\min}$, i.e., 
$U(y_{\min};\beta_{\mathrm{BP}}^{(\mathrm{SC})})
=U(\tilde{y}_{\min};\beta_{\mathrm{BP}}^{(\mathrm{SC})})$.

\section{System Models} \label{sec3}
\subsection{Spatially-Coupled CDMA Systems}  
We consider a $K$-user spatially-coupled CDMA system with 
variable spreading factor. Let $N_{t}$ and $L$ denote the spreading factor 
in symbol period~$t$ and the number of symbol periods per transmission 
block, respectively. 
The vector $\boldsymbol{u}_{k,t}\in\mathbb{C}^{N_{t}}$ transmitted by 
user~$k$ in symbol period~$t$ is given by 
\begin{equation} \label{transmitted_vector} 
\boldsymbol{u}_{k,t} = \frac{1}{\sqrt{N_{t}}}\sum_{l=0}^{L-1}h_{t,l}
\boldsymbol{s}_{t,l}^{(k)}b_{k,l} 
\quad \hbox{for $t=0,\ldots,L-1$,}  
\end{equation}
where $b_{k,l}\in\mathbb{C}$ denotes the $l$th data symbol for user~$k$ 
with unit power $\mathbb{E}[|b_{k,l}|^{2}]=1$; 
$\boldsymbol{s}_{t,l}^{(k)}\in\mathbb{C}^{N_{t}}$ represents  
the $l$th spreading sequence for user~$k$ in symbol period~$t$ with 
$\mathbb{E}[\|\boldsymbol{s}_{t,l}^{(k)}\|^{2}]=N_{t}$; and 
$h_{t,l}\in\mathbb{C}$ denotes the $l$th (deterministic) coupling 
coefficient in symbol period~$t$. 
For notational convenience, we have defined the 
transmitted vector~(\ref{transmitted_vector}) as if $L$ different 
spreading sequences are used in each symbol period. However, 
the actual number of spreading sequences is much smaller 
than $L$, because many coupling coefficients are adjusted to zero, as 
shown in the following examples.  
Note that the identical data symbols $\boldsymbol{b}_{k}=
(b_{k,0},\ldots,b_{k,L-1})^{\mathrm{T}}$ are 
transmitted in all symbol periods, while different data symbols are sent 
in the conventional multi-code CDMA systems. 
Under the assumption of unfaded channels, the received vector 
$\boldsymbol{y}_{t}\in\mathbb{C}^{N_{t}}$ in symbol period~$t$ is given by 
\begin{equation} \label{CDMA} 
\boldsymbol{y}_{t} = \sum_{k=1}^{K}\boldsymbol{u}_{k,t} + \boldsymbol{w}_{t},  
\quad \boldsymbol{w}_{t}\sim\mathcal{CN}(\boldsymbol{0},\sigma^{2}
\boldsymbol{I}_{N_{t}}). 
\end{equation}
It is straightforward to extend our analysis to the case of fading 
channels~\cite{Takeuchi082}.  

For the simplicity of analysis, the data symbols $\{b_{k,l}\}$ are assumed 
to be independent and identically distributed (i.i.d.) for all $k$ and $l$. 
Furthermore, we assume that the spreading sequences 
$\{\boldsymbol{s}_{t,l}^{(k)}\}$ are i.i.d.\ for all $k$, $t$, and $l$, and 
that each spreading sequence $\boldsymbol{s}_{t,l}^{(k)}$ have i.i.d.\ 
zero-mean real and imaginary parts with variance $1/2$. 
In order to restrict the average transmit power 
$\mathbb{E}[\|\boldsymbol{u}_{k,t}\|^{2}]$ to unit power, we impose the 
constraint $\sum_{l=0}^{L-1}|h_{t,l}|^{2}=1$. Furthermore, we impose 
the constraint $\sum_{t=0}^{L-1}|h_{t,l}|^{2}=1$ to equalize the power 
used for transmission of each data symbol. 

\begin{example}[Uncoupled System] \label{ex1} 
Let $h_{t,l}=\delta_{t,l}$ and $N_{t}=N$. The system reduces to the 
conventional CDMA system. The sum rate for quadrature phase shift keying 
(QPSK) is given by $2\beta$, with $\beta=K/N$ denoting the system load. 
\end{example}
\begin{example}[Circularly-Coupled Systems] \label{ex2} 
The coupling coefficients for circularly-coupled systems are given by 
$h_{t,l}=h_{t-l}$ for $t-l\geq 0$ and $h_{t,l}=h_{t-l+L}$ for 
$t-l<0$, with 
\begin{equation}
h_{t} = \left\{
\begin{array}{ll}
1/\sqrt{W+1} & \hbox{for $t=0,\ldots,W$} \\ 
0 & \hbox{for $t=W+1,\ldots,L-1$,} 
\end{array}
\right.
\end{equation} 
where $W+1$ corresponds to the number per symbol period of spreading sequences 
used by each user. 
One transmission block consists of an initialization phase 
$t=0,\ldots,W-1$ and a communication phase $t=W,\ldots,L-1$.  
The spreading factor $N_{t}=N_{\mathrm{init}}$ in the initialization phase 
is adjusted to a small value, so that the receiver should be able to detect 
the data symbols sent in this phase successfully. The spreading factor 
$N_{t}=N$ in the communication phase is chosen to a large value in order to 
increase the transmission rate. The sum rate $R$ for the circularly-coupled 
systems with QPSK is given by 
\begin{IEEEeqnarray}{rl}  
R &= \frac{2KL}{N_{\mathrm{init}}W + N(L-W)} 
\nonumber \\ 
&= \frac{2}{\beta_{\mathrm{init}}^{-1}(W/L) + \beta^{-1}\{1-(W/L)\}}, 
\label{sum_rate}
\end{IEEEeqnarray} 
where $\beta_{\mathrm{init}}=K/N_{\mathrm{init}}$ and $\beta=K/N$ denote the 
system loads in the initialization and communication phases, respectively. 
The system load $\beta$ in the communication phase corresponds to the 
parameter $\beta$ in the dynamical system~(\ref{dynamical_system}).  
The sum rate~(\ref{sum_rate}) tends to the sum rate $2\beta$ for the uncoupled 
CDMA system presented in Example~\ref{ex1} when $L$ tends to infinity 
with $W$ fixed.  
In other words, the rate loss due to the initialization phase is negligible 
when $L$ is sufficiently large. 
\end{example}

\subsection{IO Receiver} \label{sec3_2} 
There is an interesting connection between the performance of BP-based 
receivers and of the IO receiver for uncoupled CDMA systems, i.e., 
a fixed-point equation characterizing the performance of the IO receiver 
coincides with the BP fixed-point equation obtained from 
DE~\cite{Kabashima03,Montanari06,Guo08,Ikehara07}.  
Thus, we can predict the BP fixed-point equation by analyzing the performance 
of the IO receiver.  
  
The IO receiver detects the data symbols $\boldsymbol{b}_{k}$ for each user 
from the received vectors $\mathcal{Y}=\{\boldsymbol{y}_{t}\}$ in all symbol 
periods. The soft decision $\hat{b}_{k,l}$ of the data symbol 
$b_{k,l}$ for the IO receiver is given by the posterior mean estimator (PME) 
\begin{equation} \label{IO_receiver} 
\hat{b}_{k,l} = \sum_{\boldsymbol{b}_{k}} 
b_{k,l}p_{\boldsymbol{b}_{k} | \mathcal{Y}, \mathcal{S}}
(\boldsymbol{b}_{k} | \mathcal{Y}, \mathcal{S}),  
\end{equation}
where $p_{\boldsymbol{b}_{k} | \mathcal{Y}, \mathcal{S}}
(\boldsymbol{b}_{k} | \mathcal{Y}, \mathcal{S})$ denotes the 
posterior distribution of $\boldsymbol{b}_{k}$ given all received vectors 
$\mathcal{Y}$ and all spreading sequences  
$\mathcal{S}=\{\boldsymbol{s}_{t,l}^{(k)}\}$. 
It is well-known that the PME achieves the minimum mean-squared error (MMSE). 

\begin{remark}
Let us consider ``online'' detection for the circularly-coupled system with 
$W=1$, presented in Example~\ref{ex2}. The receiver detects the first data 
symbols $\{b_{k,0}\}$ and the last data symbols $\{b_{k,L-1}\}$ in the 
initialization phase. These symbols should be detected successfully 
for sufficiently small $\beta_{\mathrm{init}}$. 
In symbol period $t=1$, the receiver detects the second data symbols 
$\{b_{k,1}\}$ from the received vector $\boldsymbol{y}_{1}$. 
The known MAI due to the first data symbols is first subtracted from 
$\boldsymbol{y}_{1}$. The obtained vector can be regarded as the received 
vector in symbol period~$t=1$ for the uncoupled system. However, the average 
transmit power in this case is $1/2$, which is half the power for the 
uncoupled system. Thus, this type of online detection is suboptimal. 
\end{remark}

\section{Large-System Analysis} \label{sec4}
In order to analyze the performance of the IO receiver~(\ref{IO_receiver}), 
we consider the large-system limit in which $K$ and $\{N_{t}\}$ tend to 
infinity while $L$ and the system loads $\beta_{t}=K/N_{t}$ for all $t$ are 
fixed. We use the replica method to calculate the MSE for the IO receiver, 
following \cite{Takeuchi082}. 
The replica method is a powerful method for large-system analysis, although 
it is based on several heuristic assumptions~\cite{Tanaka02,Guo051}. 
The detailed calculation is omitted because of space limitation. 

The spatially-coupled CDMA system is decoupled into a bank of equivalent 
single-user channels in the large-system limit. This type of decoupling was 
originally shown for the conventional CDMA systems~\cite{Guo051}. 
We first define the equivalent single-user channels for user~$k$. 
The received signals for the single-user channels are given by 
\begin{equation} \label{AWGN} 
z_{t,l}^{(k)} = h_{t,l}b_{k,l} + w_{t,l}^{(k)} 
\quad \hbox{for $t=0,\ldots,L-1$,}
\end{equation}
with $w_{t,l}^{(k)}\sim\mathcal{CN}(0,\sigma_{t}^{2})$. 
The IO receiver for the single-user channels detects the data symbol 
$b_{k,l}$ from the received signals $\mathcal{Z}_{l}^{(k)}=
\{z_{t,l}^{(k)}:\hbox{for all $t$}\}$ on the basis of the posterior 
probability $p_{b_{k,l}|\mathcal{Z}_{l}^{(k)}}
(b_{k,l}|\mathcal{Z}_{l}^{(k)})$. The soft decision 
$\langle b_{k,l} \rangle$ of $b_{k,l}$ for this IO receiver is given by 
\begin{equation}
\langle b_{k,l} \rangle = \sum_{b_{k,l}}b_{k,l}
p_{b_{k,l}|\mathcal{Z}_{l}^{(k)}}(b_{k,l}|\mathcal{Z}_{l}^{(k)}). 
\end{equation}
Its MSE $\xi_{l}$ is given by 
\begin{equation} \label{MSE} 
\xi_{l} = \mathbb{E}\left[
 | b_{k,l} - \langle b_{k,l} \rangle|^{2}
\right]. 
\end{equation}
The MSE~(\ref{MSE}) depends on $\{\sigma_{t}^{2}\}$ only through the asymptotic 
signal-to-interference ratio (SIR) for the single-user channel~(\ref{AWGN}), 
given by  
\begin{equation} \label{SIR} 
\mathrm{sir}_{l} = \left(
 \sum_{t=0}^{L-1}|h_{t,l}|^{2}\sigma_{t}^{2} 
\right)^{-1}. 
\end{equation}
Thus, we can write (\ref{MSE}) as $\xi_{l}=\xi(\mathrm{sir}_{l}^{-1})$. 
The properties of MSE imply that the function $\xi(z)$ on $z\in(0,\infty)$ is 
a monotonically-increasing bounded function.  

We have not so far specified the variances $\{\sigma_{t}^{2}\}$. By defining   
the variances as the solution to coupled fixed-point equations, the 
MSE for the spatially-coupled CDMA system is connected with that for the 
decoupled systems.  

\begin{proposition} \label{proposition1} 
The MSE of the IO receiver for the spatially-coupled CDMA systems  
converges to that for the single-user channels in the large-system limit, i.e., 
\begin{equation} \label{decoupling_result} 
\mathbb{E}[|b_{k,l} - \hat{b}_{k,l}|^{2}] 
\rightarrow \xi_{l} 
\quad \hbox{for all $k$.} 
\end{equation}
In evaluating the right-hand side of (\ref{decoupling_result}), 
the variances $\{\sigma_{t}^{2}\}$ are given by the solution to the coupled 
fixed-point equations, 
\begin{equation} \label{fixed_point} 
\sigma_{t}^{2} = \sigma^{2} 
+ \beta_{t}\sum_{l'=0}^{L-1}|h_{t,l'}|^{2}\xi(\mathrm{sir}_{l'}^{-1}) 
\quad \hbox{for all $t$,}   
\end{equation}  
where $\mathrm{sir}_{l}$ is given by (\ref{SIR}). 
If the coupled fixed-point equations~(\ref{fixed_point}) have multiple 
solutions, one should choose the solution minimizing the so-called 
free energy, 
\begin{equation} \label{free_energy} 
F = \frac{1}{L}\sum_{l=0}^{L-1}I(b_{k,l};\mathcal{Z}_{l}^{(k)}) 
+ \frac{1}{L}\sum_{t=0}^{L-1}\beta_{t}^{-1}D(\sigma^{2}\|\sigma_{t}^{2}), 
\end{equation}
where $D(\sigma^{2}\|\sigma_{t}^{2})$ denotes the Kullback-Leibler divergence 
between $\mathcal{CN}(0,\sigma^{2})$ and $\mathcal{CN}(0,\sigma_{t}^{2})$. 
\end{proposition}

The free energy~(\ref{free_energy}) is a quantity proportional to 
the achievable sum rate 
$I(\{\boldsymbol{b}_{k}\};\mathcal{Y}|\mathcal{S})/(\sum_{t=0}^{L-1}N_{t})$ 
in the large-system limit. 

In order to obtain an intuitive understanding of the solution to 
(\ref{fixed_point}), we consider the circularly-coupled system with 
$\beta_{\mathrm{init}}\rightarrow0$, presented in Example~\ref{ex2}. 
It should be possible to detect the data symbols $\{b_{k,l}:l=0,\ldots,W-1,L-W,
\ldots,L-1\}$ transmitted in the initialization phase without errors. Thus, 
the corresponding boundaries $\{\xi_{l}\}$ can be regarded as zero. 
In positions distant from the boundaries, on the other hand, the coupled 
fixed-point equations~(\ref{fixed_point}) can be approximated by 
the phenomenological model~(\ref{stationary_system}):  
Substituting (\ref{fixed_point}) into (\ref{SIR}) yields  
\begin{equation}
\mathrm{sir}_{l}^{-1} = \sigma^{2} + \beta\xi_{l} 
+ \frac{\beta}{(W+1)^{2}}\sum_{t=0}^{W}\sum_{l'=0}^{W}\Delta_{t-l'}, 
\end{equation}
for $W\leq l\leq L-1-W$, with $\Delta_{k}=\xi_{l+k}-\xi_{l}$. 
Let $y(x)$ denote a smooth function on $x\in[0,1]$ satisfying 
$y(l/L)=\xi_{l}$ for $l=0,\ldots,L-1$. 
Applying the approximation in $L\rightarrow\infty$\footnote{
This approximation was also considered in \cite{Hassani10}. 
}, 
\begin{equation} \label{expansion} 
\Delta_{k} = y^{(1)}\left(
 \frac{k}{L}
\right) + \frac{y^{(2)}}{2}\left(
 \frac{k}{L}
\right)^{2} 
+ \frac{y^{(3)}}{3!}\left(
 \frac{k}{L}
\right)^{3} + O(L^{-4}), 
\end{equation}
with $y^{(i)}=d^{i}y/dx^{i}|_{x=l/L}$, we obtain (\ref{stationary_system}) with 
\begin{equation}
D = \frac{\beta}{2(W+1)^{2}L^{2}}\sum_{t=0}^{W}\sum_{l'=0}^{W}(t-l')^{2}, 
\end{equation}
\begin{equation} \label{CDMA_potential} 
U(y;\beta) = \int_{0}^{y} (\xi^{-1}(y') - \sigma^{2} - \beta y')dy', 
\end{equation} 
where $\xi^{-1}(z)$ denotes the inverse function of $\xi(z)$. 
Thus, the argument presented in Section~\ref{sec2} implies that 
the BP threshold $\beta_{\mathrm{BP}}^{(\mathrm{SC})}$ for the 
circularly-coupled CDMA systems in $L\rightarrow\infty$ is given by 
the point $\beta$ at which the potential energy~(\ref{CDMA_potential}) has 
two global stable solutions. Unfortunately, the potential energy is different 
from the free energy for the uncoupled CDMA systems. Thus, the BP threshold 
$\beta_{\mathrm{BP}}^{(\mathrm{SC})}$ does not coincide with the IO threshold 
$\beta_{\mathrm{IO}}$ for the uncoupled systems at which the free 
energy~(\ref{free_energy}) for the uncoupled systems has two global stable 
solutions.

\section{Comparison of Thresholds} \label{sec5}
The fixed-point equation characterizing the performance of the IO receiver 
for the uncoupled CDMA systems coincides with the BP fixed-point equation in 
the large-system limit, as mentioned in Section~\ref{sec3_2}. 
Let us assume that this statement also holds for 
spatially-coupled CDMA systems, i.e., the coupled fixed-point 
equations~(\ref{fixed_point}) are assumed to coincide with the BP fixed-point 
equations. Thus, the performance of BP-based receivers is equal to that 
of the IO receiver if the solution to the coupled fixed-point 
equations~(\ref{fixed_point}) is unique. Otherwise, the BP-based receivers 
are outperformed by the IO receiver. We hereafter focus on QPSK data symbols. 

\begin{figure}[t]
\begin{center}
\includegraphics[width=\hsize]{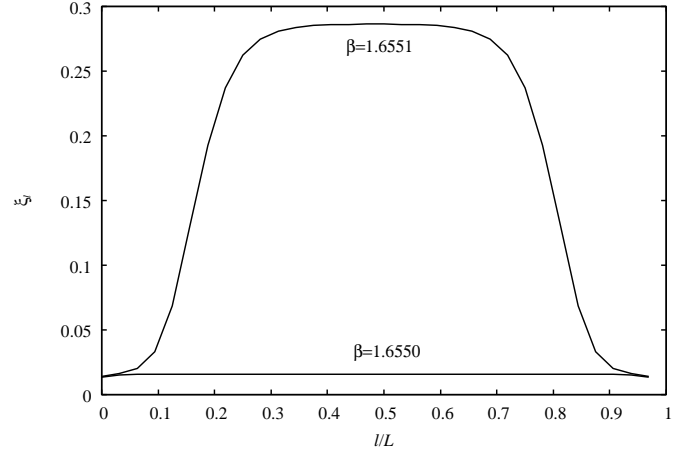}
\end{center}
\caption{
MSE for $l$th data symbol. $1/\sigma^{2}=9$~dB, $L=32$, $W=1$, 
and $\beta_{\mathrm{init}}=1.22$. 
}
\label{fig2} 
\end{figure}

We solve the coupled fixed-point equations~(\ref{fixed_point}) by successive 
iteration, i.e., 
\begin{equation} \label{time_evolution} 
\sigma_{t}^{2}(i+1) = \sigma^{2} + \beta_{t}\sum_{l'=0}^{L-1}|h_{t,l'}|^{2} 
\xi\left(
 \sum_{t'=0}^{L-1}|h_{t',l'}|^{2}\sigma_{t'}^{2}(i)
\right), 
\end{equation}
with $\sigma_{t}^{2}(0)=\sigma_{\mathrm{init}}^{2}$. 
We conjecture from the result for the 
uncoupled CDMA systems~\cite{Guo08} that (\ref{time_evolution}) with 
$\sigma_{\mathrm{init}}^{2}=\infty$ corresponds to the DE 
equation for a BP-based receiver in iteration~$i$.    
Figure~\ref{fig2} shows the MSEs calculated from the stationary solutions to 
(\ref{time_evolution}) with $\sigma_{\mathrm{init}}^{2}=\infty$. 
When $\beta=1.6550$, the MSEs take a small value for all $l$. 
Furthermore, we confirmed that they coincide with those obtained from the 
stationary solution to (\ref{time_evolution}) with 
$\sigma_{\mathrm{init}}^{2}=0$. Thus, the coupled 
fixed-point equations~(\ref{fixed_point}) has the unique solution when 
$\beta=1.6550$. When $\beta$ increases slightly, however, a 
spatially-nonuniform stationary solution appears. This stationary solution 
does not coincide with that for $\sigma_{\mathrm{init}}^{2}=0$. 
The criterion based 
on the free energy~(\ref{free_energy}) implies that the spatially-nonuniform 
solution is not the solution corresponding to the IO receiver. Thus, the 
BP-based receivers are outperformed by the IO receiver when $\beta$ is 
strictly larger than the BP threshold 
$\hat{\beta}_{\mathrm{BP}}^{(\mathrm{SC})}=1.6550$. This 
numerically-evaluated threshold 
$\hat{\beta}_{\mathrm{BP}}^{(\mathrm{SC})}=1.6550$ coincides with the 
theoretical prediction $\beta_{\mathrm{BP}}^{(\mathrm{SC})}=1.6550$ based on 
the effective potential energy~(\ref{CDMA_potential}).  
It is worth noting that we chose a large system load 
$\beta_{\mathrm{init}}=1.22$ in the initialization phase. 
One need not adjust the system load in the initialization 
phase to zero. 

\begin{table}[t]
\begin{center}
\caption{Numerically-evaluated BP threshold 
$\hat{\beta}_{\mathrm{BP}}^{(\mathrm{SC})}$ 
for circularly-coupled systems. $1/\sigma^{2}=10$~dB 
and $\beta_{\mathrm{init}}=0$.}
\label{table1} 
\begin{tabular}{|c|c|c|c|c|c|c|}
\hline
\multicolumn{2}{|c|}{} & \multicolumn{5}{|c|}{$W$} \\ 
\cline{3-7} 
\multicolumn{2}{|c|}{} & 0 & 1 & 2 & 3 & 4 \\ 
\hline 
 & 16 & 1.7307 & 1.8123 & 1.8241 & 1.8684 & 1.9455 \\ 
\cline{2-7} 
$L$ & 32 & 1.7307 & 1.8120 & 1.8121 & 1.8130 & 1.8179 \\ 
\cline{2-7} 
 & 64 & 1.7307 & 1.8120 & 1.8121 & 1.8121 & 1.8121 \\ 
\hline
\end{tabular}
\end{center}
\end{table}

Table~\ref{table1} lists the numerically-evaluated BP thresholds 
$\hat{\beta}_{\mathrm{BP}}^{(\mathrm{SC})}$ based on (\ref{time_evolution}) 
for several $L$ and $W$. 
The thresholds for $W=0$ correspond to the BP threshold for the uncoupled 
CDMA systems. The BP thresholds for the circularly-coupled CDMA systems are 
always larger than that for the uncoupled CDMA systems. The thresholds 
in right-upper cells on Table~\ref{table1} are above the theoretical 
prediction $\beta_{\mathrm{BP}}^{(\mathrm{SC})}=1.8121$ based on the potential 
energy~(\ref{CDMA_potential}). This observation is because the rate loss 
due to the initialization phase is not negligible in the right-upper region.  
Table~\ref{table1} implies that $W=1$ is the best option in terms of the sum 
rate~(\ref{sum_rate}). Note that $W$ may affect the convergence property of the 
BP-based receivers.  

Table~\ref{table2} presents the comparison between several thresholds. 
The thresholds $\beta_{\mathrm{BP}}$ and $\beta_{\mathrm{IO}}$ denotes the 
BP and IO thresholds for the uncoupled CDMA systems, respectively. The 
threshold $\hat{\beta}_{\mathrm{BP}}^{(\mathrm{SC})}$ represents the 
numerically-evaluated BP threshold for the circularly-coupled CDMA systems, 
while $\beta_{\mathrm{BP}}^{(\mathrm{SC})}$ is its theoretical prediction 
based on the potential energy~(\ref{CDMA_potential}). We calculated an upper 
bound of the IO threshold $\beta_{\mathrm{IO}}^{(\mathrm{SC})}$ for the 
circularly-coupled systems from the two stationary solutions to 
(\ref{time_evolution}) with the initial values 
$\sigma_{\mathrm{init}}^{2}=\infty$ and $\sigma_{\mathrm{init}}^{2}=0$. 
Note that the coupled fixed-point equations~(\ref{fixed_point}) have the 
unique solution when the signal-to-noise ratio (SNR) $1/\sigma^{2}$ is below 
$8.25$~dB. The theoretical prediction $\beta_{\mathrm{BP}}^{(\mathrm{SC})}$ 
is in good agreement with the numerically-evaluated one 
$\hat{\beta}_{\mathrm{BP}}^{(\mathrm{SC})}$ for $1/\sigma^{2}=9$~dB and 
$1/\sigma^{2}=10$~dB. The BP threshold 
$\hat{\beta}_{\mathrm{BP}}^{(\mathrm{SC})}$ for the circularly-coupled CDMA 
systems are larger than the BP threshold $\beta_{\mathrm{BP}}$ for the 
uncoupled CDMA systems for all SNRs, while there is a gap between 
$\beta_{\mathrm{BP}}^{(\mathrm{SC})}$ and the IO threshold 
$\beta_{\mathrm{IO}}$. 


\section{Conclusions} \label{sec6} 
We have proposed the spatially-coupled CDMA systems in order to improve 
the performance of BP-based multiuser detection. 
The large-system analysis shows that the circularly-coupled CDMA systems 
can improve the BP threshold, compared to the one for the conventional 
CDMA systems, while there is a gap between the improved BP threshold and the 
IO thresholds. We conclude that spatial coupling is a general method for 
boosting the BP threshold for graphical models up to a value, although it 
depends on the models whether or not the value coincides with the optimal one. 

\section*{Acknowledgment}
The work of K.~Takeuchi was in part supported by the Grant-in-Aid for 
Research Activity Start-up (No. 21860035) from MEXT, Japan. 

\begin{table}[t]
\begin{center}
\caption{Comparison of thresholds. $L=32$, $W=1$, and 
$\beta_{\mathrm{init}}=0$.}
\label{table2} 
\begin{tabular}{|c||c|c|c|c|c|}
\hline
$1/\sigma^{2}$ & $\beta_{\mathrm{BP}}$ & 
$\hat{\beta}_{\mathrm{BP}}^{(\mathrm{SC})}$ & 
$\beta_{\mathrm{BP}}^{(\mathrm{SC})}$ & 
$\beta_{\mathrm{IO}}$ & $\beta_{\mathrm{IO}}^{(\mathrm{SC})}$ \\ 
\hline 
9~dB & 1.6147 & 1.6550 & 1.6550 & 1.6992 & 1.7048 \\ 
\hline
10~dB & 1.7307 & 1.8120 & 1.8121 & 1.9826 & 1.9873 \\
\hline
12~dB & 1.8734 & 2.0030 & 2.0039 & 2.5071 & 2.4973 \\
\hline
14~dB & 1.9552 & 2.1109 & 2.1132 & 2.9855 & 2.9584 \\
\hline
\end{tabular}
\end{center}
\end{table}

\bibliographystyle{IEEEtran}
\bibliography{IEEEabrv,kt-isit2011_1}

\begin{thebibliography}{10}
\providecommand{\url}[1]{#1}
\csname url@samestyle\endcsname
\providecommand{\newblock}{\relax}
\providecommand{\bibinfo}[2]{#2}
\providecommand{\BIBentrySTDinterwordspacing}{\spaceskip=0pt\relax}
\providecommand{\BIBentryALTinterwordstretchfactor}{4}
\providecommand{\BIBentryALTinterwordspacing}{\spaceskip=\fontdimen2\font plus
\BIBentryALTinterwordstretchfactor\fontdimen3\font minus
  \fontdimen4\font\relax}
\providecommand{\BIBforeignlanguage}[2]{{%
\expandafter\ifx\csname l@#1\endcsname\relax
\typeout{** WARNING: IEEEtran.bst: No hyphenation pattern has been}%
\typeout{** loaded for the language `#1'. Using the pattern for}%
\typeout{** the default language instead.}%
\else
\language=\csname l@#1\endcsname
\fi
#2}}
\providecommand{\BIBdecl}{\relax}
\BIBdecl

\bibitem{Felstrom99}
A.~J. Felstr\"om and K.~S. Zigangirov, ``Time-varying periodic convolutional
  codes with low-density parity-check matrix,'' \emph{IEEE Trans. Inf. Theory},
  vol.~45, no.~6, pp. 2181--2191, Sep. 1999.

\bibitem{Kudekar10}
S.~Kudekar, T.~Richardson, and R.~Urbanke, ``Threshold saturation via spatial
  coupling: Why convolutional {LDPC} ensembles perform so well over the
  {BEC},'' in \emph{Proc. 2010 IEEE Int. Symp. Inf. Theory}, Austin, TX, USA,
  Jun. 2010, pp. 684--688.

\bibitem{Lentmaier10}
M.~Lentmaier and G.~P. Fettweis, ``On the thresholds of generalized {LDPC}
  convolutional codes based on protographs,'' in \emph{Proc. 2010 IEEE Int.
  Symp. Inf. Theory}, Austin, TX, USA, Jun. 2010, pp. 709--713.

\bibitem{Takeuchi111}
K.~Takeuchi, T.~Tanaka, and T.~Kawabata, ``A phenomenological study on
  threshold improvement via spatial coupling,'' \emph{{\rm submitted to} IEICE
  Trans. Fundamentals}, 2011, [Online]. Available:
  http://arxiv.org/abs/1102.3056.

\bibitem{Verdu98}
S.~Verd\'u, \emph{Multiuser Detection}.\hskip 1em plus 0.5em minus 0.4em\relax
  New York: Cambridge University Press, 1998.

\bibitem{Tanaka02}
T.~Tanaka, ``A statistical-mechanics approach to large-system analysis of
  {CDMA} multiuser detectors,'' \emph{{IEEE} Trans. Inf. Theory}, vol.~48,
  no.~11, pp. 2888--2910, Nov. 2002.

\bibitem{Kabashima03}
Y.~Kabashima, ``A {CDMA} multiuser detection algorithm on the basis of belief
  propagation,'' \emph{J. Phys. A: Math. Gen.}, vol.~36, no.~43, pp.
  11\,111--11\,121, Oct. 2003.

\bibitem{Montanari06}
A.~Montanari and D.~N.~C. Tse, ``Analysis of belief propagation for non-linear
  problems: The example of {CDMA} (or: How to prove {Tanaka's} formula),'' in
  \emph{Proc. 2006 IEEE Inf. Theory Workshop}, Punta del Este, Uruguay, Mar.
  2006, pp. 160--164.

\bibitem{Guo08}
D.~Guo and C.-C. Wang, ``Multiuser detection of sparsely spread {CDMA},''
  \emph{{IEEE} J. Sel. Areas Commun.}, vol.~26, no.~3, pp. 421--431, Apr. 2008.

\bibitem{Takeda06}
K.~Takeda, S.~Uda, and Y.~Kabashima, ``Analysis of {CDMA} systems that are
  characterized by eigenvalue spectrum,'' \emph{Europhys. Lett.}, vol.~76,
  no.~6, pp. 1193--1199, 2006.

\bibitem{Richardson08}
T.~Richardson and R.~Urbanke, \emph{Modern Coding Theory}.\hskip 1em plus 0.5em
  minus 0.4em\relax New York: Cambridge University Press, 2008.

\bibitem{Takeuchi082}
K.~Takeuchi, T.~Tanaka, and T.~Yano, ``Asymptotic analysis of general multiuser
  detectors in {MIMO DS-CDMA} channels,'' \emph{{IEEE} J. Sel. Areas Commun.},
  vol.~26, no.~3, pp. 486--496, Apr. 2008.

\bibitem{Ikehara07}
T.~Ikehara and T.~Tanaka, ``Decoupling principle in belief-propagation-based
  {CDMA} multiuser detection algorithm,'' in \emph{Proc. 2007 IEEE Int. Symp.
  Inf. Theory}, Nice, France, Jun. 2007, pp. 2081--2085.

\bibitem{Guo051}
D.~Guo and S.~Verd\'u, ``Randomly spread {CDMA}: Asymptotics via statistical
  physics,'' \emph{{IEEE} Trans. Inf. Theory}, vol.~51, no.~6, pp. 1983--2010,
  Jun. 2005.

\bibitem{Hassani10}
S.~H. Hassani, N.~Macris, and R.~Urbanke, ``Coupled graphical models and their
  thresholds,'' in \emph{Proc. 2010 IEEE Inf. Theory Workshop}, Dublin,
  Ireland, Aug.--Sep. 2010.

\end{thebibliography}

\end{document}